# Mueller Matrix Polarimetry of Fiber Bragg Grating Strain and Torsion

Hani J. Kbashi[1*], Alberto R. Cuevas[1], and Sergey Sergeyev[1*]

1 *Aston Institute of Photonic Technologies, College of Engineering and Physical Sciences, Aston University B4 7ET Birmingham, United Kingdom.*
*h.kbashi@aston.ac.uk, s.sergeyev@aston.ac.uk

**Abstract:** We experimentally demonstrate a polarimetric dual-comb spectroscopy technique for simultaneous strain and torsion sensing using a single-cavity mode-locked fiber laser and fiber Bragg grating (FBG) sensors. Dual-comb generation in a single-cavity fiber laser was achieved by utilizing a piece of high-birefringence fiber and adjusting the in-cavity polarization controller, resulting in a polarimetric temporal interferogram with a duration of 1.428 ms, corresponding to a repetition rate difference of 700 Hz. Fast Fourier Transform (FFT) analysis was applied to the time-domain Stokes parameters, enabling the detection of FBG spectral shifts induced by strain and torsion. The system exhibited linear responses to both strain and torsional inputs, with measured sensitivities of 25 Hz/με and 5.5 Hz/°, respectively, across a dynamic range of 600 με and 90°. To further enhance discrimination between strain and torsion, we applied a novel approach to extract Mueller matrix elements without using complex adjustable polarization components. We explored the analysis of polarimetric purity of the FBG's Mueller matrix in terms of polarizance, diattenuation, and structural polarization response as a function of FBG strain and torsion. The obtained results enabled the measurement of strain and torsion based on a single FBG, which paves the way for the development of cost-effective shape sensing technologies.

## 1. Introduction

The state of polarization (SOP) of light provides information on a light beam's source, along any interaction of the materials under the test [1, 2]. Given that the information can be obtained by using the non-invasive techniques, polarization measurements have a wide range of applications, including materials characterization [2–5], biomedical diagnostics [6-11], remote sensing [6, 12-15], and structural health monitoring [16-18]. For example, in remote sensing, it can differentiate artificial structures and natural surfaces [12-15], provide information on soil, sand, and volcanic ash properties, snow and ice characteristics such as age and types, as well as plants and ground [6, 12-15], while in medical diagnostics, it enhances contrast in biological tissues [6-11]. Unlike conventional sensing methods, which often rely solely on intensity or spectral information, polarimetric sensing offers an additional layer of information about surface texture, geometry, and stress distribution [2-18]. This makes it particularly effective in the structural health monitoring of engineering structures for detecting surface cracks, material degradation, and deformation in infrastructures such as bridges, buildings, pipelines, roads, offshore renewable energy facilities, and civil structures, supporting improved lifetime prediction and yield optimization [12-15].

Traditional polarimeter designs, as shown in Fig. 1, utilize linear polarizers and rotating quarter-wave plates to measure the Stokes parameters of the light reflected from the target, thereby gaining access to information about the object's texture [2]. In this design, series of laser pulses are transmitted through the polarization state generator (PSG), with each pulse exiting the PSG with a different state of polarization (SOP) due to the rotating quarter-wave plate. Likewise, in the polarization state analyzer (PSA), different return SOPs are analyzed for each pulse due to the rotating retarder. A sample's 4x4 Mueller matrix is estimated through serial measurements of the signal transmitted/scattered from the sample, with each measurement employing a different illumination SOP [2]. In the context of SHM, Mueller matrix polarimetry enables the detection of polarization-dependent optical response of FBGs by tracking transformations in Stokes space. By measuring the device's response to a set of input polarization states, the complete Mueller matrix can be reconstructed, revealing information about polarization-dependent losses (PDL [17, 18]) and depolarization effects in terms of degree of polarization (DOP [16]). Although providing many advantages, the state-of-the-art polarimetry of FBGs poses challenges in discriminating FBGs' strain and twist.

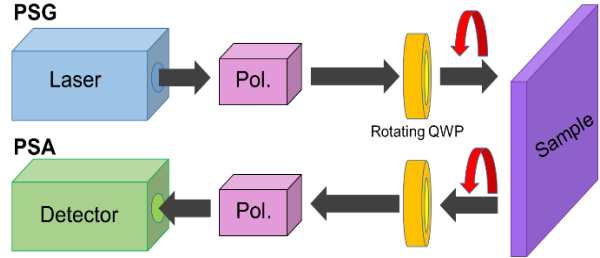

Fig. 1 Conventional dual rotating wave plate polarimeter. PSG - polarization state generator, PSA - the polarization state analyzer, Pol – polarizer, QWP - quarter-wave plate.

Dual comb spectroscopy (DCS) is one of the advanced techniques that can provide a cost-effective approach to Muller matrix polarimetry in various applications, including precision spectroscopy, distance ranging, greenhouse gas detection, biomedical imaging, and infrastructure sensing [19-33]. The DCS technique utilizes two coherent optical combs (slow and fast) with repetition rates of $f_r$ and $f_r+\Delta f$, respectively. The interaction (beating) between these optical combs generated a third radiofrequency (RF) comb, with spacing determined by $\Delta f$, establishing a one-to-one mapping between the optical and RF combs, scaled by a coefficient factor of $f_r/\Delta f$ [19-33]. Recently emerged DCS based on a single laser is a cost-effective technique that does not require complex electronic stabilization and supports long-term stability of $\Delta f$ [25-33]. The application of DCS for interrogating fiber Bragg gratings (FBGs) provides an enhanced sensing platform by combining the precision and spectral resolution of DCS with the spatial multiplexing and robustness of FBGs, utilizing only low-bandwidth electronics [25-30].

Unlike the previous study of Mueller matrix polarimetry [16-18] and DCS applications for FBGs-based sensing [25-30], in this paper, for the first time, we demonstrate experimentally a novel Muller matrix polarimetric technique based on a single cavity dual-comb source. The dual-combs were generated from a single fiber laser cavity enabling common noise cancellation, mutual coherent properties and without need of complex phase-locking subsystems that are required for two independent laser sources. By mapping the maxima of four sequential beatings of dual comb in terms of four sets of Stokes parameters, we substitute application of polarization state generator and polarization state analyzer by more effective technique that does not require additional adjustable polarization components. Applying FFT to the Stokes parameters, we found that the frequency increased linearly with applying strain and torsion to the FBG with the sensitivities of 25 Hz/με and 5.5 Hz/°, respectively, over a strain range of 600 με and a rotation range of 90°. Mueller Matrix elements have been extracted from reference and reflected sets of Stokes parameters to analyze the FBG's polarimetric purity in terms of such component's depolarization index, polarizance, diattenuation and the polarization response parameters as a function of strain and torsion. The results can enable application of a single FBG for simultaneous measurements of strain and torsion.

## 2. Experimental Setup

Fig. 2 illustrates the experimental setup for polarimetric dual–comb fiber laser spectroscopy. The dual comb has been designed and developed based on the polarization multiplexing technique in a single-ring cavity fiber laser [31-33]. The ring cavity consists of 0.45m of high-concentration Er-doped fiber (ER110-4/125) that is pumped using a 980 nm laser diode through 980/1550 wavelength division multiplexing (WDM). A composite single-wall carbon nanotube film (CNT) is used for mode-locking and ultrashort pulse generation, while a 51 dB dual-stage polarization-independent optical isolator (OISO) is used in the laser cavity to ensure a unidirectional propagation. The polarization controllers POC1 and POC2 are used to adjust the pump and cavity SOP, respectively. To enable dual-comb lasing, a 0.5 m of polarization-maintaining fiber (PMF) is inserted inside the laser cavity, which introduces significant group-velocity mismatch for pulses of different polarization states due to the slightly different refractive indices ($n_x$, $n_y$) of the slow and fast polarization modes in the PMF. Hence, a dual-comb is generated with a slight difference in repetition rates for the proof-of-concept of the dual-comb demonstration. Polarization-resolved measurements were extensively studied in our previous work by extracting the dual-comb pulses from the cavity and splitting them into two orthogonal components using a polarization beam splitter (PBS) [31, 33]. The stability of the combs' repetition rates in free running was investigated in [31, 33]. The two combs can be separated with a crosstalk suppression of more than 20 dB by adjusting POC2, suggesting that the polarization states of the two combs were nearly orthogonal. Finally, a 90:10 output optical coupler is used to redirect 90% back to the cavity and 10% outside the cavity, making the final cavity length of 5.25 m. The dual comb output is amplified from 0.5 mW to 3 mW using an EDFA and launched to the FBG sensor via port 1 of the optical circulator. The FBG served as the sensing element that responded to strain and torsion deformations by shifting the spectrum of the reflected light. The reflected dual comb from the FBG (port 2) is directed to port 3 and then split into four optical paths to monitor and analyze the results. The first path is connected directly to the OSA to monitor the optical spectrum, while the second path is launched to the fast polarimeter (NOVOPTEL 100MHz) that has a built-in low-pass filter to analyze the Stokes parameters in the time domain. The obtained dynamic waveforms have been further processed by using the Fast Fourier Transform FFT to reconstruct the FBG spectrum in the radiofrequency domain. The optical signals in paths three and four are converted to electrical signals through photodiodes (PD1 and PD2) to characterize the dual comb in temporal and RF domains using an oscilloscope (Agilent DSOX93204A) and a 13 GHz RF spectrum analyzer (Rohde and Schwarz).

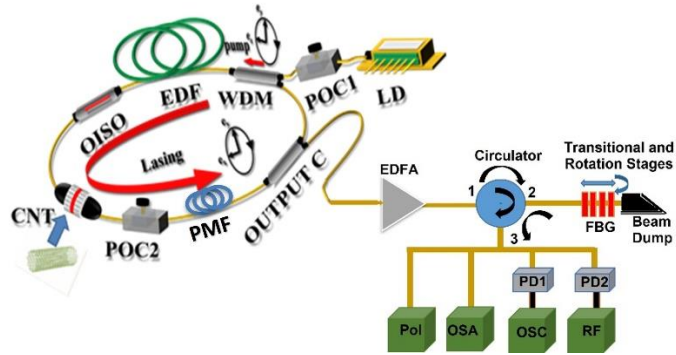

Fig. 2. Dual comb fiber laser strain sensing spectroscopy: WDM: wavelength division multiplexing, EDF: erbium doped fiber, OISO: optical isolator CNT: carbon nanotubes-based saturable absorber, POC: polarization controllers, PMF: polarization maintaining fiber, FBG: fiber Bragg grating, PD: photodiode, OSA: optical spectrum analyzer, OSC: oscilloscope, RF: radio frequency spectrum analyzer.

## 3. Results and Discussions

By adjusting the pump power and POC2 inside the laser cavity to the appropriate position, the fiber laser can be switched from stable mode-locked to operate in a simultaneous two-comb (fast and slow) regime centered at fundamental repetition rates of 39.246516 MHz and 39.247171 MHz, respectively. The corresponding round-trip times for the fast and slow modes are 25.479968718 ns and 25.479543427 ns, respectively, with a time delay of 0.425 ps. A noticeable tuning in the repetition rate difference of the two combs was observed by adjusting the pump power and POC2 [33].

The repetition rate difference (beating note) of 700 Hz that is measured by a photodetector and an RF spectrum analyzer is achieved by controlling the cavity birefringence. In the time domain, the beating produces spikes with a period of 1.428 ms. The proposed fiber laser interrogates FBG with Bragg wavelengths of 1559.5 nm. As shown in Fig. 3 (a), the reflected dual-comb spectrum without applying strain and torsion to the FBG was successfully reconstructed in the RF domain with clear two resolved frequency peaks and beating notes distributed in groups across 100 kHz of bandwidth. The SNR of both slow and fast combs is 70 dB and 60 dB, respectively. This generally reflects the high mutual coherence supported by evidence of comb beating note resolved RF spectrum, where the linewidth of the comb teeth in the RF domain is much smaller than the comb beat note tooth spacing (700 Hz), developing a phase stable dual–comb interferogram. The resulting beat notes in the form of a multi-peak structure are formed due to the interference between the two combs, which leads to a periodic beating interferogram of 1.428 ms measured by a photodetector and oscilloscope, as shown in Fig. 3 (b). The interpolate in the Fig. 3 (b) illustrates the detailed picture of the periodic beating interferogram formation inside the laser cavity, where each line of the slow comb beats with the corresponding lines of the fast comb conducting multi-heterodyne detection.

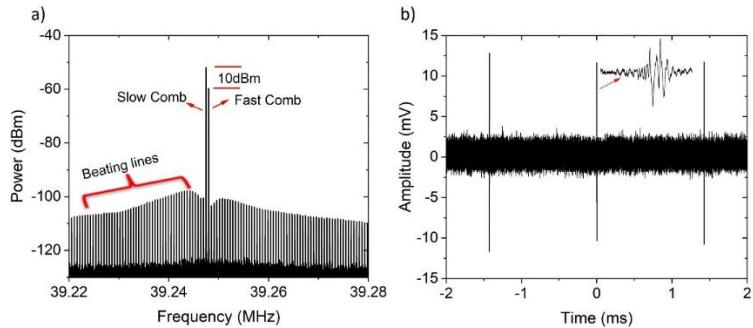

Fig. 3: (a) RF dual comb beat note and (b) temporal interferogram traces of the reflected dual comb signals without applying strain and torsion to the FBG.

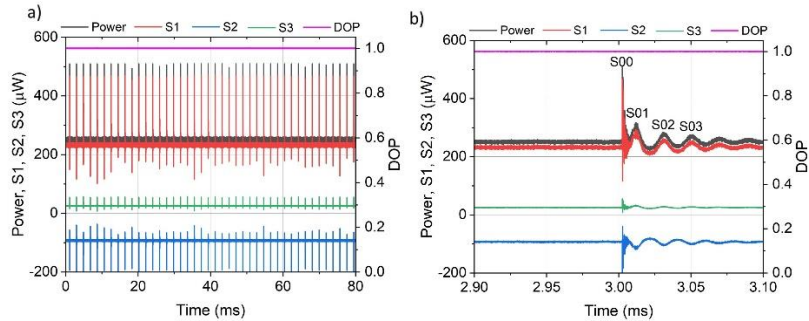

Fig. 4: (a) Spatiotemporal intensity evolution of the dual comb interferogram patten, (b) Stokes parameters interferogram trace of the reflected dual comb signals without applying strain and torsion to the FBG.

In the polarimetric domain, the sampling interferograms are generated due to the optical field interference (beating) between the slow and fast dual comb pulses resolved through down-converting heterodyne detection when the waveform reflected from the FBG is measured without applying strain to the FBG. A fast polarimeter (PM1000 Novotel, 100 MHz bandwidth) is used to record all Stokes parameters ($S_1$, $S_2$, $S_3$) as well as the power ($S_0$) and the degree of polarization (DOP). As we can see from Fig. 4(a), the Stokes parameters and DOP, the heterodyne detection produces an equally spaced multiple interferogram peaks with a separation of 1.428 ms that is measured by the fast parameters at an accomplished single-shot acquisition of 80 ms with a resolution of 80 ns (12.5MHz). The zoom of the interferogram peak shown in inset of Fig. 4(b) shows an oscillating pattern with a good signal-to-noise ratio. The mutual coherence and phase stability of the dual comb are also investigated in the polarimeter results by representing the Stokes parameters ($S_1$-$S_3$).

The polarimetric results of power are analyzed by using FFT to obtain the frequency component of the dynamic strain applied to the FBG from the interferogram around a peak. The FFT time-frequency analysis provides the depth-resolved dual comb spectroscopy result, as the strain and torsion reflected information in different portions of the interferogram trace, corresponding to different depth positions that can be retrieved. However, by taking the FFT of just one interferogram and fitting the spectral shape to the normalized Gaussian shape, the obtained reflectivity spectra of the FBGs were successfully reconstructed as shown in Fig. 5 (a), which verified the feasibility of the free-running fiber laser for DCS measurement of FBG spectrum. The Stokes signals observed in the real-time scale and presented in the FFT is compressed by the factor $f_r/\Delta f_r$ is 56KHz in our case, for the separation between the adjacent interferograms of 1.428 ms ($1/\Delta f_r$). Fig. 5 (b) illustrates the averaging FFT for the interferogram across 80 ms of the S0 temporal domain. From the rapprochement, even without any averaging, the measured FBG spectrum shows good signal quality for accurate determination of its spectral position, and this illustrates the quality of our single cavity dual-comb laser.

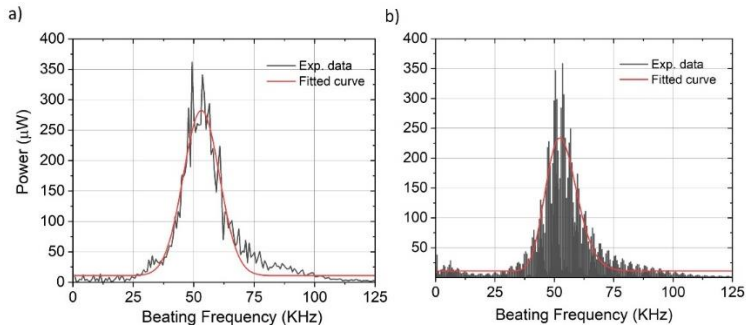

Fig. 5: (a) FFT for one interferogram pulse of $S_0$ and (b) FFT for all interferogram pulses of $S_0$.

Next, the FBG is used as the sample under test and examined for strain and torsion sensing in both the optical and temporal domains. Strains ranging from 0 to 600 µε were applied to the FBG using a digital translation stage with a length of 20 cm between the FBG points and step increments of 15 µm. Similar to the traditional strain sensors, the FBG sensor also exhibited torsion-sensitive properties where the FBG was rotated from 0° to 90° in 5° increments using a rotation stage in a positive (clockwise) torsion.

In the optical domain, the resultant strain and torsion sensitivity are found to be 1.37 pm/µε and 3 pm/° due to the variations of FBG resonance wavelength with applied strain and torsion. By using DCS technique, it was found that, the beating frequency of the reflected dual comb is increased linearly with applied strain and the FBG rotation angle as shown in Fig 6(a) and Fig. 6(b), exhibiting sensitivities of 25 Hz/µε and 0.95 Hz/°, respectively, over a strain range of 600 µε and a rotation range of 90°. When strain and torsion are applied to the FBG, the grating period and effective refractive index are altered due to strain-optic and geometric effects, resulting in a spectral shift that can be reconstructed from the shot-by-shot interferogram in the temporal domain.

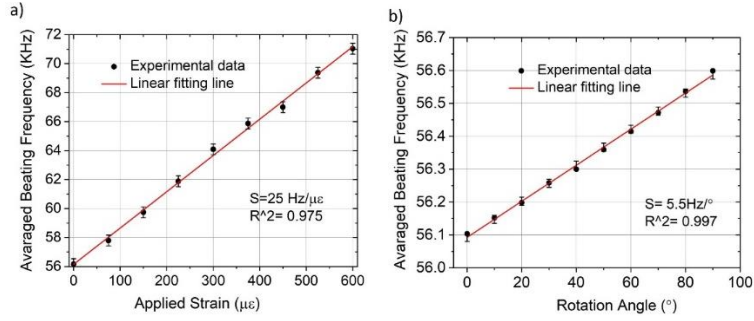

Fig. 6: (a) averaged frequency peak shift vs applied strain; (b) averaged frequency peak shift vs FBG rotation angle.

One of the primary challenges in FBG-based sensing is distinguishing the spectral peak shifts caused by axial strain from those induced by torsional (twist) effects. In this work, this challenge is addressed through analyzing the depolarization characteristics of the FBG response, explicitly using the depolarization index ($P_d$) and its associated polarization purity components [2, 34, 35]. These intrinsic polarization metrics provide a macroscopic view of the FBG's depolarizing behavior under the combined influence of strain and twist.

The depolarization index generalizes the concept of the DOP and serves as a quantitative measure of a FBG depolarizing power. So, it is a Mueller matrix that quantifies the optical system scrambles or reduces the DOP of reflected light from the FBG. The purity components comprise a set of three parameters that establish a direct connection between depolarization characteristics and the dichroic properties of the FBG. These parameters include the polarizance (P), representing the magnitude of the polarizance vector that is derived from the first column of the Mueller matrix; the diattenuation (D) is the Mueller matrix that describes optical system preferentially reflects different polarization states of light, So, it represents the magnitude of the polarization-dependent attenuation of light (diattenuation vector); and a third parameter, $P_s$, which further characterizes the structural polarization response by the non-depolarizing part of the Mueller matrix [2, 34, 35]. These parameters are derived from the elements of the $4 \times 4$ Mueller matrix (M), which quantifies the transformation of polarization states by the medium. The M is constructed by measuring the absolute transmitted or reflected intensities for various input polarization states, and the purity metrics are then computed using established relationships from the matrix elements.

$$\boldsymbol{S}_T = \mathbf{M}_T \cdot \boldsymbol{S}_R. \qquad (1)$$

Where $S_R$ and $S_T$ are the experimentally observed quantities of the reference and reflected Stokes vectors from the FBG at different strain and rotation angles. It is represented in form of

$$\boldsymbol{S}_T = \begin{bmatrix} S_{T00} & S_{T01} & S_{T02} & S_{T03} \\ S_{T10} & S_{T11} & S_{T12} & S_{T13} \\ S_{T20} & S_{T21} & S_{T22} & S_{T23} \\ S_{T30} & S_{T31} & S_{T32} & S_{T33} \end{bmatrix}, \mathbf{M}_T = \begin{bmatrix} m_{00} & m_{01} & m_{02} & m_{03} \\ m_{10} & m_{11} & m_{12} & m_{13} \\ m_{20} & m_{21} & m_{22} & m_{23} \\ m_{30} & m_{31} & m_{32} & m_{33} \end{bmatrix},$$

$$S_R = \begin{bmatrix} S_{R00} & S_{R01} & S_{R02} & S_{R03} \\ S_{R10} & S_{R11} & S_{R12} & S_{1R3} \\ S_{R20} & S_{R21} & S_{R22} & S_{R23} \\ S_{R30} & S_{R31} & S_{R32} & S_{R33} \end{bmatrix}.$$

(2)

Where $S_{R00}$, $M_{00}$, $S_{T00}$ are the mean intensity transmittance and $S_{Rij}$, $M_{ij}$, $S_{Tij}$ are the element at the *i*-th row and *j*-th column of $S_T$, $M$, and $S_R$.

Muller matrix $M$ can be expressed in the following partitioned form [2, 33, 34]:

$$M = m_{00} \begin{pmatrix} 1 & D^T \\ P & m \end{pmatrix},$$

(3)

Where

$$P \equiv \frac{1}{m_{00}}(m_{10}, m_{20}, m_{30})^T, D \equiv \frac{1}{m_{00}}(m_{01}, m_{02}, m_{03})^T, m \equiv \frac{1}{m_{00}}\begin{pmatrix} m_{11} & m_{12} & m_{13} \\ m_{21} & m_{22} & m_{23} \\ m_{31} & m_{32} & m_{33} \end{pmatrix}.$$

(4)

Here $D$ and $P$ are the diattenuation and polarizance vectors. The lengths of these vectors are diattenuation $D = |D|$ and polarizance $P = |P|$ [2, 33, 34]. Thus, the mean transmittance for incoming unpolarized light for matrix $M$ is $m_{00}$, whereas the degree of polarimetric purity of is mapped by the depolarization index $P_d$ [2, 33, 34]:

$$P_d = \sqrt{D^2 + P^2 + |m|_2^2},$$

(5)

where $|m|_2$ is Euclidean norm of the submatrix $m$ [2, 33, 34].

Based on Eqs. (3)-(5), the $D, P, |m|_2$ and depolarization index $P_d$ can be found as follows [2, 33, 34]:

$$D = \sqrt{\frac{\sum_{j=1}^{3} m_{0j}^2}{m_{00}}}, P = \sqrt{\frac{\sum_{i=1}^{3} m_{i0}^2}{m_{00}}}, P_s = |m|_2 = \sqrt{\frac{\sum_{i,j=1}^{3} m_{ij}^2}{3m_{00}}}, P_d^2 = \frac{1}{3}P^2 + \frac{1}{3}D^2 + P_s^2.$$

(6)

For nondepolarizing Mueller matrix $P_d = 1$ and nonpure or depolarizing Mueller matrix, $P_d < 1$ [2, 33, 34].

By analyzing the polarization parameters under varying levels of applied strain and twist, it was observed that the values of P and D exhibited a minimal linear increase variation with increasing axial strain applied to the FBG. At the same time, P is decreased linearly, yielding almost a constant $P_d$ values across the applied strain. The slight increase in the D and Ps values is likely attributed to a slight reduction in the fiber core diameter under tensile loading, which in turn may reduce the fiber's intrinsic birefringence, an effect primarily captured by the P parameter.

In contrast, when a torsional load was applied (i.e., FBG rotation), the core geometry remained largely unchanged. As a result, the parameters D, P$_s$, and P$_d$ showed a slightly larger linear increasing variation, while a vast decline in the P, indicating that twist-induced effects on the FBG's polarization response are comparatively subtle. The large observed fluctuations in these values are attributed to the localized modifications in the fiber's birefringent properties caused by mechanical twisting.

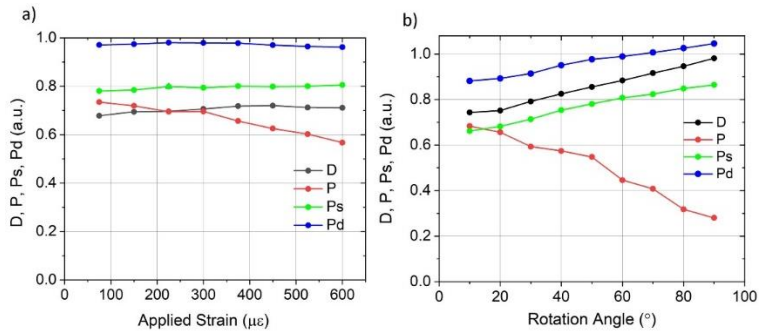

Fig. 7: components of purity and depolarization index parameters at (a) different FBG applied strain; (b) different FBG rotation angle.

## 4. Conclusions

By exploring a single-cavity dual-comb source based on a polarization multiplexing technique, we experimentally demonstrate a novel cost-polarimetric approach for simultaneous fiber Bragg grating strain and torsion sensing with measured sensitivities of 25 Hz/µε and 5.5 Hz/°, respectively, across a dynamic range of 600 µε and 90°. By measuring the maxima of four sequential beatings of dual comb in terms of four sets of Stokes parameters, we extract data on Mueller matrix elements without using additional adjustable polarization components. The analysis of polarimetric purity of the FBG's Mueller matrix shows different responses of polarizance, diattenuation, and structural polarization to applied strain and torsion. The obtained results can enable the application of a single FBG for simultaneous measurements of strain and torsion, which paves the way for the development of cost-effective shape sensing technologies.

**Acknowledgment.** Authors acknowledge support from the EPSRC project EP/W002868/1, the IF RAENG IF2223B-133 and the Royal Society IF\R1\241042.

**Disclosures.** The authors declare no conflicts of interest.

**Data availability.** Data underlying the results presented in this paper are not publicly available at this time but may be obtained from the authors upon reasonable request.

## References

1. E. Wolf, "Introduction to the Theory of Coherence and Polarization of Light," Cambridge University, (2007).
2. J. J. Gil and R. Ossikovski, "Polarized Light and the Mueller Matrix Approach," CRC Press, (2016).
3. D. Stifter, P. Burgholzer, O. Höglinger, *et al.*, "Polarization-sensitive optical coherence tomography for material characterization and strain-field mapping," Appl. Phys. A **76**(6), 947-951 (2003).
4. Z. Li, C. Cui, X. Zhou, *et al.*, "Characterization of amorphous carbon films from 5 nm to 200 nm on single-side polished a-plane sapphire substrates by spectroscopic ellipsometry," Front. Phys. **10**, 968101 (2022).
5. H. Fujiwara, "Spectroscopic Ellipsometry: Principles and Applications," John Wiley & Sons, (2007).
6. D. N. Ignatenko, A. V. Shkirin, Y. P. Lobachevsky, *et al.*, "Applications of mueller matrix polarimetry to biological and agricultural diagnostics: A review," Appl. Sci. **12**(10), 5258 (2022).
7. C. He, H. He, J. Chang, *et al.*, "Polarization optics for biomedical and clinical applications: a review," Light: Sci. Appl. **10**(1), 194 (2021).
8. N. Ghosh and I. A. Vitkin, "Tissue polarimetry: concepts, challenges, applications, and outlook," J. Biomed. Opt. **16**(11), 110801 (2011)
9. J. C. Ramella-Roman, I. Saytashev, and M. Piccini, "A review of polarization-based imaging technologies for clinical and preclinical applications," J. Opt. **22**(12), 123001 (2020).
10. I. Pardo, S. Bian, J. Gomis-Brescó, *et al.*, "Wide-field mueller matrix polarimetry for spectral characterization of basic biological tissues: Muscle, fat, connective tissue, and skin," J. Biophotonics **17**(1), e202300252 (2024).

11. V. V. Tuchin, "Polarized light interaction with tissues," J. Biomed. Opt. **21**(7), 071114 (2016).
12. X. Liu, L. Zhang, X. Zhai, *et al.*, "Polarization lidar: Principles and applications," Photonics **10**(10), 1118 (2023).
13. Z. Kong, T. Ma, Y. Cheng, *et al.*, "A polarization-sensitive imaging lidar for atmospheric remote sensing," J. Quant. Spectrosc. Radiat. Transf. **1**(271), 107747 (2021).
14. Y. Han, D. Salido-Monzú, and A. Wieser, "Classification of material and surface roughness using polarimetric multispectral LiDAR," Opt. Eng., **62**(11), 114104-114104 (2023).
15. D. Yuan, and D. Pan, "Estimation of the primary productivity and particulate organic carbon in inland waters using shipborne polarimetric lidar," Optics Express, **33**(3), 6048-6069 (2025).
16. A.W. Domanski, M. Bieda, P. Lesiak, *et al.*, "Polarimetric Optical Fiber Sensors for Dynamic Strain Measurement in Composite Materials," Acta Physica Polonica A, **124** (3), (2013).
17. C. Caucheteur, T. Guo, and J. Albert, "Polarization-assisted fiber Bragg grating sensors: Tutorial and review," J. Light. Technol., **35**(16), 3311-3322 (2016).
18. O. Xu, S. Lu, and S. Jian, "Theoretical analysis of polarization properties for tilted fiber Bragg gratings," Science China Information Sciences, **53**(2), 390-397 (2010).
19. I. Coddington, N. Newbury, and W. Swann, "Dual-comb spectroscopy," Optica **3**, 414-426 (2016).
20. B. Xu, Z. Chen, T. Hänsch, W. N. Picqué, "Near-ultraviolet photon-counting dual-comb spectroscopy," Nature **627**, 289–294 (2024).
21. T-A Liu, N. R. Newbury, and I. Coddington, "Sub-micron absolute distance measurements in sub-millisecond times with dual free-running femtosecond Er fiber-lasers," Opt. Express **19**, 18501-18509 (2011).
22. I. Coddington, W. Swann, L. Nenadovic, *et al.*, "Rapid and precise absolute distance measurements at long range," Nature Photon **3**, 351–356 (2009).
23. J. Chen, X. Zhao, Z. Yao, *et al*., "Dual-comb spectroscopy of methane based on a free-running Erbium-doped fiber laser," Optics Express, 27(8), pp. 11406-11412 (2019).
24. F. R. Giorgetta, J. Peischl, D. I. Herman, *et al.*, "Open-Path Dual-Comb Spectroscopy for Multispecies Trace Gas Detection in the 4.5–5 μm Spectral Region," Laser & Photonics Reviews **15**, 2000583 (2021).
25. J. Guo, K. Zhao, B. Zhou, *et al.*, "Wearable and Skin-Mountable Fiber-Optic Strain Sensors Interrogated by a Free-Running Dual-Comb Fiber Laser," Advanced Optical Materials **7**, 1900086 (2019).
26. J. Guo Y. Ding, X.Xiao, *et al.*, "Multiplexed static FBG strain sensors by dual-comb spectroscopy with a free running fiber laser," Optics Express, **26** (13), 16147-16154 (2018).
27. R. Zhang, Z. Zhu, G. Wu, "Static pure strain sensing using dual–comb spectroscopy with FBG sensors," Opt. Express. **27** (23), 34269-34283 (2019).
28. N. Kuse, A. Ozawa, and Y. Kobayashi, "Static FBG strain sensor with high resolution and large dynamic range by dual-comb spectroscopy," Opt. Express, **21** (9), 11141-1149 (2013).
29. X. Zhao, J. Yang, J. Liu, *et al.*, "Dynamic Quasi-Distributed Ultraweak Fiber Bragg Grating Array Sensing Enabled by Depth-Resolved Dual-Comb Spectroscopy," IEEE Trans. Instrum. Meas., **69**, 8, 5821-5827 (2020).
30. X. Zhao, Q. Li, S. Yin, J. Chen and Z. Zheng, "Dual-Comb Dynamic Interrogation of Fiber Bragg Grating with One Mode-Locked Fiber Laser," IEEE Sensors Journal, **18**(16) 6621-6626 (2018).
31. R. Cuevas, H. J. Kbashi, D. Stoliarov, S. Sergeyev, "Polarization dynamics, stability and tunability of a dual-comb polarization-multiplexing ring-cavity fiber laser," Results in Physics, **46**, 106260, (2023).
32. A. R. Cuevas, I. Kudelin, I., H. Kbashi, S. Sergeyev, "Single-shot dynamics of dual-comb generation in a polarization-multiplexing fiber laser," Sci. Rep. **13**, 19673 (2023).
33. A. R. Cuevas, D. Stoliarov, H. Kbashi, S. Sergeyev, "Dual-comb polarization-multiplexing ring-cavity fiber laser for ranging applications," Opt. Express, **33**(8), 18161-18169 (2025).
34. A. V. Eeckhout, J. J. Gil, E. Garcia-Caurel, *et al.*, "Unraveling the physical information of depolarizers," Opt. Express **29**, 38811-38823 (2021).
35. J. J. Gil, "Review on Mueller matrix algebra for the analysis of polarimetric measurements," Journal of Applied Remote Sensing, **8**(1), 081599-081599 (2014).